\documentclass[12pt]{article}
\usepackage{amssymb}
\usepackage{amsmath}

\newtheorem{theorem}{Theorem}

\newtheorem{corollary}{Corollary}

\begin{document}

\author{{Roustam Zalaletdinov} \\ [5mm]
\emph{Department of Mathematics and Statistics, Dalhousie University} \\
\emph{Chase Building, Halifax, Nova Scotia, Canada B3H 3J5} \\ [2mm]
\emph{Department of Theoretical Physics, Institute of Nuclear Physics} \\
\emph{Uzbek Academy of Sciences, Ulugbek, Tashkent 702132, Uzbekistan, CIS} \\
}
\title{{\LARGE \textbf{Averaging out Inhomogeneous Newtonian Cosmologies:
II. Newtonian Cosmology and the Navier-Stokes-Poisson Equations }}}
\date{}
\maketitle

\begin{abstract}
The basic concepts and equations of Newtonian Cosmology are presented in the
form necessary for the derivation and analysis of the averaged
Navier-Stokes-Poisson equations. A particular attention is paid to the
physical and cosmological hypotheses about the structure of Newtonian
universes. The system of the Navier-Stokes-Poisson equations governing the
cosmological dynamics of Newtonian universes is presented and discussed. A
reformulation of the Navier-Stokes-Poisson equations in terms of the fluid
kinematic quantities is given and the structure of this system of equations
is analyzed.
\end{abstract}

\section{Introduction}

In order to derive a system of the averaged Navier-Stokes-Poisson equations
and to make use of the system for analysis of inhomogeneous Newtonian
universes, it is significant to present the basic concepts and equations of
Newtonian Cosmology in the form necessary to approach this problem. In this
presentation a particular attention will be paid to the physical and
cosmological hypotheses about the structure of Newtonian universes which are
spatially infinite configurations of the self-gravitating compressible
Newtonian cosmological fluid without viscosity and heat transfer. The system
of the Navier-Stokes-Poisson equations governing the cosmological dynamics
of Newtonian universes is presented and discussed. A reformulation of the
Navier-Stokes-Poisson equations in terms of the fluid kinematic quantities
is given and the structure of this system of equations is analyzed.

The structure of this paper is as follows. Chapter 2 describes the full set
of the Navier-Stokes-Poisson equations for the self-gravitating compressible
Newtonian fluid without viscosity and heat transfer under the gravitational
field of its own Newtonian gravitational potential. The definition of the
Newtonian-space and analysis of the coordinate freedom allowed by the
Kinematical group of the Navier-Stokes-Poisson equations is carried out in
Chapter 3. It is pointed out, in particular, that the Newtonian
gravitational potential is not uniquely defined for spatially unbounded
fluid configurations. Chapter 4 is devoted to the discussion of the initial
boundary value problem for the Navier-Stokes-Poisson equations which is not
well-posed and the Heckmann-Sch\"{u}cking boundary conditions for
homogeneous and isotropic fluid configurations are formulated. The main
hypotheses of Newtonian cosmology are formulated in Chapter 5. In Chapter 6
the Newtonian cosmological principle is formulated and the class of the
homogeneous and isotropic Newtonian universes is studied. A definition of
inhomogeneous and/or anisotropic Newtonian universes is given in Chapter 7.
The nonexistence of the static homogeneous and isotropic Newtonian universe
with the vanishing Newtonian cosmological constant is proved in Chapter 8.
Chapter 9 makes a brief comparison of Newtonian and general relativistic
cosmologies and gives evidence in favor of the remarkable physical
similarity of the both cosmological pictures. The full system of the
Navier-Stokes-Poisson equations in terms of kinematic quantities is derived
in Chapter 10. The next Chapter gives a detailed discussion of this system
of equations. It is pointed out, in particular, that in order to become
closed it requires an additional evolution equation for the tidal force
tensor which is constructed from the gravitational potential and is an
analogue of the space-time curvature tensor of general relativity. It means
that the cosmological fluid configurations governed by the
Navier-Stokes-Poisson equations in terms of kinematic quantities do evolve
due to a generalized Newtonian gravitational force. In the last Chapter 12
the Raychaudhuri evolution equation of Newtonian cosmology is shown to be
equivalent to the Friedmann equation for the cosmological scale factor for
homogeneous and isotropic Newtonian universes.

The formulae and Sections from the paper I \cite{Zala-AIC1:2002} will be
referred to as (I-XX) and I-X, correspondingly.

Conventions and notations are as follows. All functions $f=f(x^{i},t)$ are
defined on 3-dimensional Euclidean space $E^{3}$ in the Cartesian
coordinates $\{x^{i}\}$ with Latin space indices $i,j,k,...$ running from 1
to 3, and $t$ is the time variable. The Levi-Civita symbol $\varepsilon
_{ijk}$ is defined as $\varepsilon _{123}=+1$ and $\varepsilon ^{123}=+1$
and $\delta ^{ij}$, $\delta _{j}^{i}$ and $\delta _{ij}$ are the Kronecker
symbols. The symmetrization of indices of a tensor $
T_{jk}^{i}=T_{jk}^{i}(x^{l},t)$ is denoted by round brackets, $T_{(jk)}^{i}=
\frac{1}{2}\left( T_{jk}^{i}+T_{kj}^{i}\right) $, and the antisymmetrization
by square brackets, $T_{[jk]}^{i}=\frac{1}{2}\left(
T_{jk}^{i}-T_{kj}^{i}\right) $. A partial derivative of $T_{jk}^{i}$ with
respect to a spatial coordinate $x^{i}$ or time $t$ is denoted either by
comma, or by the standard calculus notation, $T_{jk,t}^{i}=\partial
T_{jk}^{i}/\partial t$ and $T_{jk,l}^{i}=\partial T_{jk}^{i}/\partial x^{l}$
, the fluid velocity is $u^{i}=u^{i}(x^{j},t)$,\ and the material (total)
derivative is $\dot{T}_{jk}^{i}\equiv dT_{jk}^{i}/dt=\partial
T_{jk}^{i}/\partial t+u^{l}\partial T_{jk}^{i}/\partial
x^{l}=T_{jk,t}^{i}+u^{l}T_{jk,l}^{i}$. The Newton gravitational constant,
the velocity of light and the Newtonian cosmological constant are $G$, $c$
and $\Lambda $, respectively.

\section{The Navier-Stokes-Poisson Equations}

\label{*nspe}

Let us consider the self-gravitating compressible Newtonian fluid without
viscosity and heat transfer under the gravitational field of its own
Newtonian gravitational potential $\phi =\phi (x^{i},t)$. The motion of the
fluid is determined by the fluid velocity field $u^{i}=u^{i}(x^{j},t)$
satisfying the compressibility condition (I-18),
\begin{equation}
\frac{\partial u^{i}}{\partial x^{i}}\neq 0,  \label{compressible}
\end{equation}
everywhere and by the fluid density field $\rho =\rho (x^{i},t)$ positive
everywhere (I-29),
\begin{equation}
\rho (x^{i},t)>0.  \label{mas-positive}
\end{equation}
The symmetric fluid stress tensor $T^{ij}=T^{ij}(x^{k},t)$ (I-40) satisfying
the Boltzmann postulate (I-75), $T^{ij}=T^{ji}$, is assumed to be the fluid
stress tensor of the perfect fluid, that is, the Newtonian fluid (I-83) with
the vanishing viscosity coefficients, $\lambda (x^{i},t)=0$ and $\mu
(x^{i},t)=0$, the vanishing heat flux vector $q^{i}(x^{j},t)=0$ and the
constant temperature $T=\mathrm{const}$, (I-79) such that
\begin{equation}
T^{ij}=-\delta ^{ij}p.  \label{perfect-fluid}
\end{equation}
An equation of state relating the fluid pressure $p=p(x^{i},t)$ and the
fluid density $\rho =\rho (x^{i},t)$ is assumed to be defined (I-81) as
\vspace{0.2cm}

\noindent \textbf{The equation of state }
\begin{equation}
p=p(\rho )\hspace{0.4cm}\mathrm{or}\hspace{0.4cm}\rho =\rho (p).
\label{eq-state2}
\end{equation}
\vspace{0.2cm}

The external force $F^{i}=F^{i}(x^{j},t)$ for the self-gravitating fluid is
defined to be due to the Newtonian gravitational potential $\phi =\phi
(x^{i},t)$,
\begin{equation}
F^{i}=-\delta ^{ij}g_{i},  \label{grav-acceleration}
\end{equation}
where $g_{i}=g_{i}(x^{j},t)$ is the Newtonian gravitational acceleration
vector of a fluid particle moving with the velocity $u^{i}(x^{j},t)$. The
acceleration vector has the following property.\vspace{0.2cm}

\noindent \textbf{The identity for the Newtonian gravitational acceleration }
$g_{i}$,
\begin{equation}
\varepsilon ^{ijk}g_{j,k}=0,  \label{identity}
\end{equation}
\vspace{0.2cm}

\noindent which means that the Newtonian gravitational field is always locally
represented by its scalar potential $\phi (x^{i},t)$
\begin{equation}
g_{i}=\phi _{,i}.  \label{potential}
\end{equation}
The Newtonian gravitational acceleration vector $g_{i}(x^{j},t)$ and the
Newtonian gravitational potential $\phi (x^{i},t)$ satisfy the Poisson
equation with the Newtonian cosmological constant $\Lambda $\vspace{0.2cm}

\noindent \textbf{The Poisson equation for the Newtonian gravitational
potential }$\phi $
\begin{equation}
\delta ^{ij}g_{i,j}=4\pi G\rho -\Lambda ,\hspace{0.4cm}\mathrm{or}\hspace{
0.4cm}\delta ^{ij}\phi _{,ij}=4\pi G\rho -\Lambda .  \label{poisson}
\end{equation}
\vspace{0.2cm}

\noindent The conservation of fluid mass (I-30)\ brings the equation of continuity
(I-31) for the fluid density $\rho (x^{i},t)$\vspace{0.2cm}

\noindent \textbf{The equation of continuity for the fluid density }$\rho $
\textbf{\ }
\begin{equation}
\rho _{,t}+(\rho u^{i})_{,i}=0,  \label{conservation}
\end{equation}
\vspace{0.2cm}

\noindent and the conservation of linear momentum (I-37) brings the Cauchy equation of
motion (I-41) which for the fluid stress tensor (\ref{perfect-fluid}) and
the external force (\ref{grav-acceleration}) becomes\footnote{
Though the equation of motion (\ref{navier-stokes}) is the Euler equation
for the perfect fluid (I-80), it will be called here the Navier-Stokes
equation of motion in accordance with Eq. (I-83) where the viscosity
coefficients are assumed to vanish and the external force is due to the
Newtonian gravitational potential (\ref{grav-acceleration}).}\vspace{0.2cm}

\noindent \textbf{The Navier-Stokes equation of motion for the fluid
velocity }$u^{i}$
\begin{equation}
u_{i,t}+u_{i,j}u^{j}=-g_{j}-\frac{1}{\rho }p_{,j}\hspace{0.4cm}\mathrm{or}
\hspace{0.4cm}u_{i,t}+u_{i,j}u^{j}=-\phi _{,i}-\frac{1}{\rho }p_{,j}.
\label{navier-stokes}
\end{equation}
\vspace{0.2cm}

\noindent Another form of the Navier-Stokes equation (\ref{navier-stokes}) can be
written by using the total fluid acceleration vector $A^{i}=A^{i}(x^{j},t)$
(I-45) for a fluid particle,
\begin{equation}
A_{i}=\frac{du_{i}}{dt}+g_{i}\equiv a_{i}+g_{i}.
\label{acceleration-total-newton}
\end{equation}
The total fluid acceleration vector $A^{i}(x^{j},t)$ can be defined (I-46)
by the Navier-Stokes equation of motion (\ref{navier-stokes}) as\vspace{0.2cm
}

\noindent \textbf{The total fluid acceleration }$A_{i}$\textbf{\ as
determined by the Navier-Stokes equation}\textsc{\ }
\begin{equation}
A_{i}=-\frac{1}{\rho }p_{,i}.  \label{navier-stokes-total}
\end{equation}
\vspace{0.2cm}

This form of the Navier-Stokes equation means from the physical point of
view that the total acceleration $A_{i}$ of a fluid particle due to the
combined effects of the inertial and gravitational forces represented by the
fluid particle's acceleration $a_{i}$ and the Newtonian gravitational
acceleration vector $g_{i}$ is determined by the gradient of the fluid
pressure. Due to the Navier-Stokes equation of motion (\ref
{navier-stokes-total}) the total fluid particle acceleration (\ref
{acceleration-total-newton}) is always directed away from a region with
higher pressure towards a region with lower pressure.

The system of seven first order partial differential equations (\ref
{identity}), (\ref{poisson}), (\ref{conservation}) and (\ref{navier-stokes})
has seven unknowns $u^{i}(x^{j},t)$, $\rho (x^{i},t)$ and $g_{i}(x^{j},t)$
with a given equation of state (\ref{eq-state2}). It is equivalent to a
system of five first and second order equations (\ref{poisson}), (\ref
{conservation}) and (\ref{navier-stokes}) for five unknowns $u^{i}(x^{j},t)$
, $\rho (x^{i},t)$ and $\phi (x^{i},t)$. This system is of a mixed
hyperbolic-elliptic type and in order to be solved in a physical setting of
interest it should be supplemented by the corresponding initial and boundary
conditions for the unknowns. As is has been discussed in Section I-13 the
problem of solving the equations of classical hydrodynamics is very
difficult. The Poisson equation brings additional difficulties when one is
looking for a solution for a self-gravitating fluid configuration satisfying
the system of the Navier-Stokes-Poisson (\ref{eq-state2}), (\ref{poisson}),
(\ref{conservation}) and (\ref{navier-stokes}). A few results have been
proved rigorously for compact Friedmann-like solutions and their finite
perturbations for the polytropic equation of state, in particular, the
linearization stability, a local-in-time existence of the solutions given a
set of data not precisely equivalent to initial data, the occurrence of
kinematical singularities (see \cite{Maki:1986}-\cite{Brau:1998} and
reference therein). Conditions for the equilibrium of the self-gravitating
rotating systems for particular classes of the equations of state have been
also studied (see \cite{FRS:2002} and references therein).

\section{The Newtonian Space-time and the Kinematical Group}

\label{*nstkg}

The Navier-Stokes-Poisson equations (\ref{poisson}), (\ref{conservation})
and (\ref{navier-stokes}) have been formulated in the framework of the field
description of fluid motion when all objects characterizing the fluid and
its properties are defined with respect to a position $\{x^{i}\}$ at a time $
t$, see Section I-5. Before analyzing and solving the equations one must
analyze the mechanics of the fluid motion and reveal the freedom in the
choice of possible Eulerian coordinate systems, which is available as far as
the equations are formulated in this framework.

As it has been pointed out in Section I-9, the Cauchy equation of motion
(I-41) is the Newton second law for a moving fluid particle. From the point
of view of Newtonian mechanics the dynamics of the whole fluid configuration
under consideration can be represented in the field description by a
sequence of its space configurations given as 3-dimensional spaces
represented by the full set of all fluid particle's positions $\{x^{i}\}$
which are occupied at each instant of time $t$. The time $t$ at positions $
\{x^{i}\}$ is assumed to be measured by a family of ideal clocks each of
which indicates values of time regardless of the previous motion. This is
the concept of absolute time. On the basis of this picture of the fluid
motion one can naturally define the notion of a Newtonian space-time manifold
\footnote{
The concept of a Newtonian space-time is known to have led to a geometic
formulation of Newtonian gravity in the framework of the pseudo-Riemanian
geometry \cite{Trau:1965}, \cite{Cart:1923}, \cite{Cart:1924}.} as follows
\cite{Trau:1965}, \cite{Stew:1990}.\vspace{0.2cm}

\noindent \textbf{The Newtonian space-time\hspace{0.2cm}}\emph{A Newtonian
space-time is defined as a 4-dimensional differentiable manifold of the
following structure:}

\emph{(a) the real-valued absolute time }$t$\emph{\ is determined up to a
linear transformation }$t\rightarrow at+b$, $t,a,b\in R$, $t>0$, $a>0$;

\emph{(b) the 3-dimensional spaces }$t=\mathrm{const}$\emph{\ are Euclidean
spaces }$E^{3}$\emph{\ endowed with a coordinate system given by the
Eulerian space coordinates }$\{x^{i}\}$;

\emph{(c) for each value of time }$t$\emph{\ there is only one 3-dimensional
space }$E^{3}$.\vspace{0.2cm}

Let us consider now the Cauchy equation of motion (I-41) in the absence of
all external and internal forces,
\begin{equation}
\rho \frac{du^{i}}{dt}=0\hspace{0.4cm}\mathrm{as}\hspace{0.4cm}F^{i}=0,
\hspace{0.4cm}T^{ij}=0.  \label{cauchy-free}
\end{equation}
which corresponds the state of free motion of each fluid particle. The
following theorem takes place \cite{Trau:1965}.

\begin{theorem}[The inertial coordinates and the Galilean group]
If a fluid is in free motion (\ref{cauchy-free}) there always exists a
global system of the Eulerian coordinates $\{x^{i},t\}$ called the inertial
coordinates which are defined up to the Galilean group of arbitrary linear
transformation,
\begin{equation}
x^{i}\rightarrow A_{j}^{i}x^{j}+B^{i}t+C^{i},\hspace{0.4cm}t\rightarrow at+b,
\label{galilean}
\end{equation}
where $A_{j}^{i}$, $B^{i}$ and $C^{i}$ are real and constant, and the matrix
$A_{j}^{i}$ is nonsingular.
\end{theorem}

\noindent \textbf{Proof.}\hspace{0.4cm}The group of Galilean transformations
(\ref{galilean}) is found as the general solution of the Cauchy equation for
free motion (\ref{cauchy-free}) for the Eulerian space coordinates $
\{x^{i}\} $ along fluid particle paths (I-2), $x^{i}=x^{i}(t)$,
\begin{equation}
\frac{d^{2}x^{i}}{dt^{2}}=0,  \label{cauchy-free-path}
\end{equation}
and the law of allowed transformations of the absolute time given above.
\hspace{0.4cm}\textbf{QED}\vspace{0.2cm}

Evidently the paths of fluid particles moving freely (\ref{cauchy-free}) are
straight lines in the Newtonian space-time of the fluid.

In the presence of gravitation when a self-gravitating fluid moves in its
own Newtonian potential $\phi =\phi (x^{k},t)$ in accordance with the Cauchy
equation of motion (I-41) with the external force (\ref{grav-acceleration})
and the fluid stress tensor $T^{ij}=0$,
\begin{equation}
\frac{du^{i}}{dt}=-\delta ^{ij}\frac{\partial \phi }{\partial x^{j}}\hspace{
0.4cm}\mathrm{as}\hspace{0.4cm}F^{i}=-\delta ^{ij}g_{i},\hspace{0.4cm}
T^{ij}=0.  \label{cauchy-newton}
\end{equation}
one cannot determine a state of free motion in the sense (\ref{cauchy-free})
and (\ref{cauchy-free-path}) and define the class of inertial coordinates
(\ref{galilean}). It is natural, however, to define another class of
privileged coordinates which corresponds to the state of free fall (\ref
{cauchy-newton}) of fluid particles in the gravitational field. It should be
pointed out here that the definition of free fall assumes that the inertial
and gravitational masses of any fluid particle are the same so that given
the same gravitational field any particles move along the same paths if
their initial positions and velocities were the same. The following theorem
determines the corresponding class of coordinates and its transformation
group \cite{Trau:1965}, \cite{Stew:1990}-\cite{Heck-Schu:1956}.

\begin{theorem}[The free fall coordinates and the Kinematical group]
If a fluid is in free fall motion (\ref{cauchy-newton}) there always exists
a global system of the Eulerian coordinates $\{x^{i},t\}$ called the free
fall coordinates which are defined up to the Kinematical group of arbitrary
linear transformation,
\begin{equation}
x^{i}\rightarrow A_{j}^{i}x^{j}+D^{i}(t),\hspace{0.4cm}t\rightarrow at+b,
\label{kinematical}
\end{equation}
where $A_{j}^{i}$ is the real-valued nonsingular matrix $A_{j}^{i}$ and $
D^{i}(t)$ is a real-valued arbitrary function of time $t$. Under the change
of coordinates (\ref{kinematical}) the gravitational potential $\phi
(x^{k},t)$ undergoes the transformation
\begin{equation}
\phi \rightarrow \phi -\delta _{ij}x^{i}\frac{d^{2}D^{j}}{dt^{2}}.
\label{kinematical-pot}
\end{equation}
\end{theorem}

\noindent \textbf{Proof.}\hspace{0.4cm}The group of kinematical
transformations (\ref{kinematical}) and transformations of the potential
(\ref{kinematical-pot}) are found as the general solution of the Cauchy
equation for free fall motion (\ref{cauchy-newton}) for the Eulerian space
coordinates $\{x^{i}\}$ of a fluid particle path (\ref{path}), $
x^{i}=x^{i}(t)$,
\begin{equation}
\frac{d^{2}x^{i}}{dt^{2}}=-\delta ^{ij}\frac{\partial \phi }{\partial x^{j}},
\label{cauchy-free-fall-path}
\end{equation}
and the law of allowed transformations of the absolute time given above.
\hspace{0.4cm}\textbf{QED}\vspace{0.2cm}

It should be pointed out here that the form of the Cauchy equation of motion
(\ref{cauchy-newton}) remains the same in any free fall coordinates (\ref
{kinematical}) and (\ref{kinematical-pot}).

Thus, in the presence of gravitation the coordinates of a position $
\{x^{i}\} $ of a free falling fluid particle are defined up to the
time-independent rotation $x^{i}\rightarrow A_{j}^{i}x^{j}$ at a given
instant of time $t$, and up to\ the time-dependent translation $
x^{i}\rightarrow x^{i}+D^{i}(t)$ possibly different for different moments of
time $t$. The Newtonian potential is not a coordinate-free function, but
rather defined in dependence of a particular choice of the Eulerian
coordinates $\{x^{i},t\}$ and it changes due to (\ref{kinematical-pot}) upon
their change. There is, however, a particular class of the self-gravitating
fluid configurations called the isolated fluid configurations when a fluid
occupies a bounded compact region with a unique common Newtonian potential
\cite{Trau:1965}, \cite{Heck-Schu:1955}, \cite{Heck-Schu:1956}.

\begin{corollary}[The isolated fluid configurations]
The Newtonian gravitational potential $\phi $ (\ref{kinematical-pot}) for
the isolated fluid configurations satisfying the global boundary condition
at spatial infinity
\begin{equation}
\phi (x^{k},t)\rightarrow 0\hspace{0.4cm}\mathrm{as}\hspace{0.4cm}\left(
\delta _{ij}x^{i}x^{j}\right) ^{1/2}\rightarrow \infty
\label{boundary-isolated}
\end{equation}
is uniquely defined,
\begin{equation}
\phi \rightarrow \phi .  \label{isolated-pot}
\end{equation}
Then the Kinematical group (\ref{kinematical}) reduces to the Galilean group
(\ref{galilean}),
\begin{equation}
D^{j}(t)=B^{i}t+C^{i}.  \label{kinematical-isolated}
\end{equation}
\end{corollary}

For the case of isolated fluid configurations (\ref{boundary-isolated}) one
is able to distinguish in an invariant manner the inertial and gravitational
parts of the total acceleration $A^{i}$ (\ref{acceleration-total-newton}).
Indeed, the left-hand side of the Cauchy equation of motion for the free
fall (\ref{cauchy-free-fall-path}) which is an inertial acceleration $a^{i}$
and its right-hand side which is a Newtonian gravitational acceleration $
g^{i}$ are kept invariant independently under the Kinematical group (\ref
{kinematical}) with (\ref{kinematical-isolated}).

\section{The Initial Boundary Value Problem}

\label{*ibvp}

In general case of the noncompact fluid configurations when the
self-gravitating fluid and therefore the gravitational field are extended
all over a noncompact 3-dimensional Euclidean space $E^{3}$ such a boundary
condition at spatial infinity (\ref{boundary-isolated}) cannot be imposed.
Such cosmological fluid configurations are considered in modelling Newtonian
universes represented by Newtonian space-times filled with the matter in the
form of self-gravitating cosmological fluid distributions over in noncompact
3-dimensional Euclidean spaces $E^{3}$ evolving during an interval of time
possibly infinite. In those cases one is forced to make a choice either to
abandon the concept of the inertial Eulerian coordinates (\ref{galilean}) or
to announce the free fall coordinate systems (\ref{kinematical}) inertial.
The choice is usually made in favour of the latter \cite{Trau:1965}, \cite
{Heck-Schu:1955}-\cite{Bond:1960}. From the physical point of view, such a
choice means that all observers belonging the same 3-dimensional Euclidean
space $E^{3}$ at $t=\mathrm{const}$ are inertial though such inertial
observers may accelerate relatively to each other when they move with the
fluid due its gravitation.

The initial boundary value problem for the system of the
Navier-Stokes-Poisson equations (\ref{eq-state2}), (\ref{poisson}), (\ref
{conservation}) and (\ref{navier-stokes}) is known not to be well-posed. A
3-dimensional Euclidean space $t=\mathrm{const}$ is a characteristic surface
of the system and it is not permitted to to set initial data on such a
surface \cite{Cour-Hilb:1966}. As a result, the potential is known up to an
arbitrary harmonic function $\psi =\psi (x^{i},t)$ satisfying a boundary
value problem for the Laplace equation $\delta ^{ij}\psi _{,ij}=0$ since it
is not defined uniquely (\ref{kinematical-pot}). Now with the partial time
derivatives of the fluid density $\partial \rho /\partial t$ and the fluid
velocity $\partial u^{i}/\partial t$ being determined from the equation of
continuity (\ref{conservation}) and the Navier-Stokes equation of motion
(\ref{navier-stokes}), the time derivative of the potential $\partial \phi
/\partial t$ must satisfy a boundary value problem for the Poisson equation
(\ref{poisson}) without $\Lambda $ term,
\begin{equation}
\delta ^{ij}\frac{\partial ^{2}}{\partial x^{i}\partial x^{j}}\frac{\partial
\phi }{\partial t}=4\pi G\frac{\partial \rho }{\partial t}.
\label{possion-time-der}
\end{equation}
which gives rise again to an arbitrary harmonic function $\psi _{1}=\psi
_{1}(x^{i},t)$ satisfying a boundary value problem for the Laplace equation $
\delta ^{ij}(\psi _{1,t})_{,ij}=0$ due to (\ref{kinematical-pot}). Such a
process eventually leads to an infinite number of boundary value problems.

A physical manifestation of the problem with the bad-posedness of the
initial value problem for the Navier-Stokes-Poisson equations can be
understood if one considers the simplest possible noncompact fluid
configuration, that is, a self-gravitating fluid with constant density, $
\rho =\rho (x^{i},t)=\mathrm{const}$. On the basis of physical arguments one
expects a homogeneous distribution of the self-gravitating fluid throughout
the all 3-dimensional spaces $t=\mathrm{const}$, since a uniform, constant
density does not show any particular point in the fluid distribution as
preferable. As there is no preferable point in the fluid distribution, it is
expected to move as a whole and there should be a particular inertial frame
where the fluid is at the state of rest at a time $t=t_{0}$. A natural
initial condition for the fluid velocity therefore is $u^{i}(x^{j},t_{0})=0$
. The Navier-Stokes-Poisson equations (\ref{poisson}), (\ref{conservation})
and (\ref{navier-stokes}) in this case read
\begin{equation}
\delta ^{ij}\frac{\partial ^{2}\phi }{\partial x^{i}\partial x^{j}}=4\pi
G\rho -\Lambda ,\hspace{0.4cm}\frac{du^{i}}{dt}=-\frac{\partial \phi }{
\partial x^{i}},\hspace{0.4cm}\mathrm{for}\hspace{0.4cm}\rho (x^{i},t)=
\mathrm{const},\hspace{0.4cm}u^{i}(x^{j},t_{0})=0.  \label{density-constant}
\end{equation}
The Poisson equation in the system (\ref{density-constant}) can be solved in
the spherical coordinates $\left( r,\theta ,\varphi \right) $ and the
resulting solution the Navier-Stokes-Poisson equations
\begin{equation}
u^{i}(x^{j},t)=-\frac{1}{3}\left( 4\pi G\rho -\Lambda \right) x^{i}(t-t_{0}),
\hspace{0.4cm}\phi (x^{i})=\frac{1}{6}\left( 4\pi G\rho -\Lambda \right)
r^{2},  \label{density-constant-sol}
\end{equation}
in contradiction with the initial assumptions for the classical Poisson
equation (\ref{poisson}) when the Newtonian cosmological constant vanishes, $
\Lambda =0$. Indeed, first of all, the solution (\ref{density-constant-sol})
appears to distinguish the origin of an arbitrarily chosen coordinate system
from other points since the gravitational potential vanishes, $\phi
(x^{i})=0 $, at $r=\left( \delta _{ij}x^{i}x^{j}\right) ^{1/2}=0$. Secondly,
the fluid cannot be at rest at a 3-dimensional space $t=t_{0}=\mathrm{const}$
, but rather moves with a constant acceleration $a^{i}(x^{k})=-4\pi Grx^{i}/3
$ away from the point with $r=0$ where the gravitational potential $\phi
(x^{i})=0$.

If now one assumes the condition $4\pi G\rho =\Lambda $, the situation
improves, since in this case the solution (\ref{density-constant-sol}) shows
that $u^{i}(x^{j},t)=0$ and $\phi (x^{i})=0$ everywhere in agreement with
the physical analysis made before approaching the equations. It should be
emphasized here that the presence of the Newtonian cosmological constant $
\Lambda $ in the Poisson equation means a modification of the classical
Poisson equation of Newtonian gravity where $\Lambda =0$. This issue is
discussed later in this Section.

Therefore, any way to improve the situation with the initial boundary value
problem for the Poisson equation and, as a result, with the
Navier-Stokes-Poisson equations, to allow a proper analysis of
self-gravitating fluid configurations defined on noncompact 3-dimensional
spaces $t=\mathrm{const}$, would assume a modification of the Poisson
equation either in its structure, or in the structure of its boundary
conditions. The latter also would result in an effective change in structure
of the Poisson equation, as any boundary condition on the gravitational
potential itself $\phi (x^{k},t)$ could not improve the situation as it is
clear from the above analysis.

The attempts of the first kind are being made since a long ago, for the
unsatisfactory situation with the application of the classical Poisson
equation to unbounded systems has been realized very soon after the
discovery of this equation. In 1895 Seeliger and Neumann \cite{Seel:1895},
\cite{Neum:1896} proposed the modified Poisson equation
\begin{equation}
\delta ^{ij}\frac{\partial ^{2}\phi }{\partial x^{i}\partial x^{j}}-\chi
\phi =4\pi G\rho ,\hspace{0.4cm}\chi =\mathrm{const},
\label{neumann-selinger}
\end{equation}
which assumes a finite range $\chi ^{-1/2}$ of this gravitational force as
compared with an infinite range of the Newtonian gravitational potential.
This equation can be shown to have the constant solution for the constant
fluid density, $\rho =\rho (x^{k},t)=\mathrm{const}$,
\begin{equation}
u^{i}(x^{k},t)=0,\hspace{0.4cm}\phi (x^{k})=\frac{4\pi }{\chi }G\rho =
\mathrm{const},  \label{neumann-selinger-const}
\end{equation}
which in agreement with the expected homogeneous fluid distribution in the
state of rest once it was at rest initially. However, one can show that this
equation can not be obtained as a weak field limit of the Einstein equations
of general relativity even in the presence of the cosmological constant $
\Lambda =\lambda c^{2}$ where $\lambda $\ is the cosmological constant
entering the Einstein equations. It is the the Poisson equation (\ref
{poisson}) that follows as the weak field limit of the Einstein equations.
This limiting procedure is singular in the sense that the Einstein equations
for perfect fluids with an equation of state (\ref{eq-state2}) have a
well-posed initial boundary problem \cite{Lich:1955}, \cite{Syng:1960}, but
the Poisson equation has not. It has been assumed therefore that a
modification of the Poisson equation must be dictated by a reconsideration
of its derivation from the Einstein equations by taking into account the
post-Newtonian approximation terms which provide the weak-field correction
terms next to the Newtonian approximation. It has been shown \cite
{Szek-Rain:2000} that one can construct a generalization of the
Navier-Stokes-Poisson equations which does possess a well-defined initial
boundary problem, but those equations have very different structure and
nonlinear in the gravitational potential itself and in its products with the
fluid velocity. Another generalization \cite{Bert-Hami:1994} of the
Navier-Stokes-Poisson equations gives a system of equations very similar to
the linearized Einstein equations which provide an equation for propagation
of the gravitational potential, to permit gravitational waves and a finite
speed of gravitational interaction as compared with the infinite speed
propagating Newtonian gravitation with no gravitational radiation.

Attempts to modify the boundary conditions for the gravitational potential
have proved to be more fruitful because in this approach one can usually
assume a condition on the basis of physical arguments \cite{Elli-Duns:1994}.
First of all, it should be noted that using a simple generalization of the
boundary condition (\ref{boundary-isolated}) such as $\phi \rightarrow \phi
(t)_{\mid \infty }$ as $\left( \delta _{ij}x^{i}x^{j}\right)
^{1/2}\rightarrow \infty $ or a prescription for the value of $\lim \delta
^{kl}\phi _{,kl}$ as $\left( \delta _{ij}x^{i}x^{j}\right) ^{1/2}\rightarrow
\infty $, does not work because the potential $\phi $ diverges at spatial
infinity and $\delta ^{kl}\phi _{,kl}$ is determined by the limiting fluid
density at infinity though the Poisson equation. Indeed, the structure of
the Poisson equation (\ref{poisson}) is known to determine the divergence of
the acceleration, $\delta ^{ij}g_{i,j}=\delta ^{ij}\phi _{,ij}$, while the
so--called tidal force tensor $E_{ij}=E_{ij}(x^{k},t)$ defined as\vspace{
0.2cm}

\noindent \textbf{The tidal force tensor }$E_{ij}$ \textbf{for the Newtonian
gravitational potential }$\phi $

\begin{equation}
E_{ij}=\phi _{,ij}-\frac{1}{3}\delta _{ij}\delta ^{kl}\phi _{,kl},\hspace{
0.4cm}E_{ij}=E_{ji},\hspace{0.4cm}\delta ^{kl}E_{kl}=0,
\label{curvature-newton}
\end{equation}
\vspace{0.2cm}

\noindent is left undetermined, the tensor being an analogue of the space-time
curvature of general relativity. Therefore eight component of the trace-free
symmetric tensor $E_{ij}$ are not determined by the Poisson equation and
they are therefore pure kinematic quantities in the Newtonian gravity. The
absence of the field equation to govern the tidal force tensor $E_{ij}$ (\ref
{curvature-newton}) may be considered as a reflection of the above problems.
On the other hand, one can use the freedom in $E_{ij}$ to get a suitable
boundary conditions for particular classes of spatially noncompact
self-gravitating fluid configurations. The suitable boundary conditions that
allow one to handle spatially homogeneous fluid configurations which
necessarily have the uniform fluid density have been formulated by Heckmann
and Sch\"{u}cking \cite{Heck-Schu:1955}, \cite{Heck-Schu:1956}, \cite
{Elli-Duns:1994}.\vspace{0.2cm}

\noindent \textbf{The Heckmann-Sch\"{u}cking boundary condition for the
tidal force tensor }$E_{ij}$\textbf{\hspace{0.2cm}}\emph{The following
boundary condition is assumed to hold for spatially homogeneous and
isotro- pic fluid configurations }\textbf{\ }
\begin{equation}
E_{ij}(x^{k},t)\rightarrow 0\hspace{0.4cm}\mathrm{as}\hspace{0.4cm}\left(
\delta _{ij}x^{i}x^{j}\right) ^{1/2}\rightarrow \infty .
\label{heckmann-schucking}
\end{equation}
\vspace{0.2cm}

\noindent This boundary condition (\ref{heckmann-schucking}) enables one to describe
consistently the unbounded homogeneous and isotropic distributions of the
self-gravitating fluid. When taken as cosmological models, such cosmological
fluid configurations are Newtonian analogues \cite{Trau:1965}, \cite
{Heck-Schu:1955}-\cite{Bond:1960} of the Friedmann-Lemaitre-Robertson-Walker
(FLRW) cosmological models of general relativity \cite{Frie:1922}-\cite
{KSHM:1980}. The conditions, however, can be shown to exclude the spatially
homogeneous but anisotropic self-gravitating cosmological fluid
configurations. Such Newtonian configurations corresponds to the
Kantowski-Sachs and Bianchi spatially homogeneous but anisotropic
cosmological models of general relativity \cite{KSHM:1980}-\cite{Bian:1897}.
A generalization of the boundary condition (\ref{heckmann-schucking}) to
cover this broader class of fluid configurations is also known \cite
{Heck-Schu:1955}, \cite{Heck-Schu:1956}, \cite{Elli-Duns:1994}.\vspace{0.2cm}

\noindent \textbf{The generalized Heckmann-Sch\"{u}cking boundary condition
for the tidal force tensor }$E_{ij}$\textbf{\hspace{0.2cm}}\emph{The
following boundary condition is assumed to hold for spatially homogeneous
anisotropic fluid configurations}
\begin{equation}
E_{ij}(x^{k},t)\rightarrow E(t)_{ij}|_{\infty }\hspace{0.4cm}\mathrm{as}
\hspace{0.4cm}\left( \delta _{ij}x^{i}x^{j}\right) ^{1/2}\rightarrow \infty .
\label{heckmann-schucking2}
\end{equation}
\vspace{0.2cm}

\noindent Here $E(t)_{ij}|_{\infty }$ are arbitrary functions of time. The boundary
conditions (\ref{heckmann-schucking}) and (\ref{heckmann-schucking2}) are
considered as physically adequate for this class of cosmological fluid
configurations (see \cite{Elli-Duns:1994} for a discussion and further
references). In both cases information immediately propagates in from
infinity to determine the local physical evolution of a moving fluid. In
this approach no new equation for the propagation of the tidal force tensor $
E_{ij}$ is explicitly necessary if one remains in the framework of Newtonian
gravity, though it leads to impossibility to decide a priori which boundary
condition fits a physical setting under consideration. If such a propagation
equation for $\partial E_{ij}/\partial t$ is defined in a framework of the
generalized Newtonian gravity, then a physical setting would dictate a
particular boundary condition at infinity, to determine nonlocally through
this condition the local dynamics of a self-gravitating cosmological fluid.

\section{The Newtonian Cosmology}

\label{*nc}

Cosmology is a theory of the evolution and structure of our Universe. In
dependence on physical and mathematical hypotheses about the structure of
the Universe space-time, the content and distribution of cosmological matter
and the set of dynamical equations governing the Universe evolution one can
formulate different theoretical frameworks for derivation and analysis of
cosmological models to compare their results with observations and
experimental data (see \cite{Trau:1965}, \cite{Bond:1960}, \cite{Elli:1984}-
\cite{Elli:1973} for a discussion and references). The following hypotheses
are assumed for Newtonian universes, the cosmological models of Newtonian
Cosmology.\vspace{0.2cm}

\noindent \textbf{The Newtonian cosmological space-time hypothesis}\textsc{
\hspace{0.2cm}}\emph{Newtonian universes are assumed to have the geometry of
Newtonian space-time. }\vspace{0.2cm}

A Newtonian universe therefore always possesses an absolute time $t$. A
state of a Newtonian universe at every instant of the absolute time $t$ is
given by a 3-dimensional Euclidean space $E^{3}$ and can be therefore
described by the Eulerian coordinates $(x^{k},t)$ (see Section \ref{*nstkg}).
\vspace{0.2cm}

\noindent \textbf{The Newtonian cosmological matter hypothesis}\textsc{
\hspace{0.2cm}}\emph{A Newtonian universe is assumed to be filled with a
self-gravitating compressible Newtonian cosmological fluid without viscosity
and heat transfer under the gravitational field of its own Newtonian
gravitational potential.}\vspace{0.2cm}

Though generally a cosmological fluid can be assumed Stokean, see Section
I-13, the Newtonian cosmological fluid is taken as a physically reasonable
model for the matter in our Universe and due to its own gravitation as the
only reason of the universe evolution. The next hypothesis determines the
physical laws of fluid dynamics and its gravitation.\vspace{0.2cm}

\noindent \textbf{The Newtonian cosmological dynamics hypothesis}\textsc{
\hspace{0.2cm}}\emph{A Newtonian universe is assumed to be governed by the
Navier-Stokes-Poisson equations (\ref{eq-state2}), (\ref{poisson}), (\ref
{conservation}) and (\ref{navier-stokes}) supplemented by a set of initial
and boundary conditions for a cosmological fluid configuration under study.}
\vspace{0.2cm}

Thus, a particular cosmological fluid configuration determines a particular
Newtonian universe. A particular cosmological fluid velocity field $
u^{i}(x^{j},t)$, the cosmological fluid density distribution $\rho (x^{j},t)$
and the Newtonian gravitational potential field $\phi (x^{j},t)$ given by a
solution to the system of the Navier-Stokes-Poisson equations (\ref
{eq-state2}), (\ref{poisson}), (\ref{conservation}) and (\ref{navier-stokes}
) for a Newtonian cosmological model determine the evolution of the
Newtonian universe as that of the cosmological fluid configuration. The
fluid velocity field $u^{i}(x^{j},t)$ determines the equation of motion of
cosmological fluid particles $x^{i}=x^{i}(\xi ^{j},t)$ and their
trajectories (I-1)-(I-6) through the differential equation (I-10),
\begin{equation}
\frac{dx^{i}}{dt}=u^{i}(x^{j},t),\quad x^{i}(0)=\xi ^{i},
\label{velocity-eqs}
\end{equation}
to predict the state of the Newtonian universe for moments of the time
interval of the cosmological evolution allowed by the initial and boundary
conditions in dependence on the initial positions $x^{i}(0)=\xi ^{i}$ of the
cosmological fluid particles. With a particular distribution of the
cosmological fluid density $\rho (x^{j},t)$, the motion of the fluid
particles taken as matter constituents should be compared with the observed
motion of the real cosmological matter to conclude about the physical
adequacy of the cosmological model and interpret and/or predict the data.

\section{The Newtonian Cosmological Principle}

\label{*ncp}

The cosmological matter in our Universe is distributed in a highly
inhomogeneous manner on the scales characteristic of typical matter
condensations such as stars, galaxies, cluster of galaxies, etc. However,
for the largest scales the overall distributions of the matter structures is
assumed to be homogeneous and isotropic \cite{Bond:1960}, \cite{Elli:1984}-
\cite{Peeb:1993}. This fundamental assumption about the large-scale
structure of the real Universe is called the Cosmological Principle. Let us
formulate this Principle in the framework of Newtonian Cosmology \cite
{Trau:1965}, \cite{Bond:1960}, \cite{Miln:1934}, \cite{McCr-Miln:1934}.
\vspace{0.2cm}

\noindent \textbf{The Newtonian cosmological principle}\textsc{\hspace{0.2cm}
}\emph{For any observer moving with a cosmological fluid particle the
Newtonian universe appears to be the same at different points of space, at
different instances and at different directions. Any other observer moving
with another cosmological fluid particle at the same time observes the same
appearance of the Newtonian universe.}\vspace{0.2cm}

The Newtonian cosmological principle assumes the existence of a privileged
family of observers who, at each instant of time, observe the Newtonian
universe as homogeneous and isotropic. The distributions of the cosmological
fluid over 3-dimensional Euclidean spaces at each instant of time are
therefore homogeneous and isotropic. However, that does not mean that the
cosmological fluid is static. In general, it does evolve in time since the
structure of the Newtonian space-time permits nontrivial time shifts (\ref
{kinematical}) between the 3-dimensinal Euclidean spaces $t=\mathrm{const}$.
A fundamental theorem based on the result of McCrea and Milne \cite
{Miln:1934}, \cite{McCr-Miln:1934} establishes the law of the evolution of a
Newtonian universe satisfying the Newtonian cosmological principle \cite
{Trau:1965}, \cite{Bond:1960}. Up to a different interpretation of one term,
this equation has the same form as the Friedmann equation \cite{Frie:1922}
governing the evolution of the FLRW universes which are homogeneous and
isotropic cosmological models in the framework of general relativity \cite
{Bond:1960}, \cite{Elli:1984}, \cite{Kras:1997}, \cite{Heck-Schu:1959}.

\begin{theorem}[The homogeneous and isotropic Newtonian universes]
The evolu- \linebreak tion of a Newtonian universe satisfying the Newtonian
cosmological principle is governed by the Friedmann equation
\begin{equation}
\left( \frac{dR}{dt}\right) ^{2}=\frac{8\pi G}{3R}\rho _{0}-k+\frac{1}{3}
\Lambda R^{2}  \label{friedmann}
\end{equation}
where $R=R(t)$ is the so-called scale factor of the cosmological fluid, $k$
is the constant of integration which represents the total energy of fluid
particles, $\rho _{0}=\rho (t_{0})$ is the initial value of the cosmological
fluid density.
\end{theorem}

\noindent \textbf{Proof.}\hspace{0.4cm}Let us consider a Newtonian universe
filled with a cosmological fluid satisfying the Newtonian cosmological
principle an governed by the Navier-Stokes-Poisson equations (\ref{eq-state2}
), (\ref{poisson}), (\ref{conservation}) and (\ref{navier-stokes}). An
observer moving with a fluid particle and having set up a system of the
Eulerian coordinates $(x^{k},t)$ with the origin at the fluid particle
measures the physical properties of a fluid particle at a position $
\{x^{k}\} $ as its the velocity $u^{i}(x^{k},t)$, the density $\rho (x^{k},t)
$ and the pressure $p(x^{k},t)$. Another observer residing on another fluid
particle with another system of the Eulerian coordinates $(x^{\prime k},t)$
with the origin at that fluid particle measures the physical properties of
the same fluid particle at the same position as the first observer, to
obtain the velocity $u^{\prime i}(x^{\prime k},t)$, the density $\rho
(x^{\prime k},t)$ and the pressure $p(x^{\prime k},t)$. Dashes over the
density and the pressure have been omitted since these fluid properties are
defined independently of any observer. The Newtonian cosmological principle
demands now that $u^{\prime i}(x^{\prime k},t)$, $\rho (x^{\prime k},t)$ and
$p(x^{\prime k},t)$ should be the same functions of $x^{\prime k}$ and $t$
as $u^{i}(x^{k},t)$, $\rho (x^{k},t)$ and $p(x^{k},t)$ are functions of $
x^{k}$ and $t$,
\begin{equation}
u^{i}(x^{\prime k},t)=u^{i}(x^{k},t),\hspace{0.4cm}\rho (x^{\prime
k},t)=\rho (x^{k},t),\hspace{0.4cm}p(x^{\prime k},t)=p(x^{k},t).
\label{observers}
\end{equation}
The second and third conditions (\ref{observers}) mean that the density and
pressure should be independent of a position of a fluid particle,
\begin{equation}
\rho =\rho (t),\hspace{0.4cm}p=p(t),  \label{density-pressure}
\end{equation}
and they are functions of the absolute time $t$ only. If the relative
velocity of the fluid particle associated with the first observer with
respect to the fluid particle associated with the second observer is $
u^{i}(x^{\prime k}-x^{k},t)$, then, by taking into account the first
condition (\ref{observers}), the velocity $u^{\prime i}(x^{\prime k},t)$ of
the fluid particle at the position $\{x^{\prime k}\}$ can be represented as
\begin{equation}
u^{i}(x^{\prime k},t)=u^{i}(x^{k},t)+u^{i}(x^{\prime k}-x^{k},t)\hspace{0.4cm
}\mathrm{or}\hspace{0.4cm}u^{i}(x^{\prime k}-x^{k},t)=u^{i}(x^{\prime
k},t)-u^{i}(x^{k},t).  \label{velocity-law}
\end{equation}
The system of functional equations (\ref{velocity-law}) has a solution
\begin{equation}
u^{i}(x^{k},t)=H_{j}^{i}(t)x^{j}.  \label{velocity-homogeneous}
\end{equation}
Now by the Newtonian cosmological principle the motion of the cosmological
fluid is isotropic, that is, it must not depend on any direction, $
H_{j}^{i}(t)=H(t)\delta _{j}^{i}$, and the cosmological fluid velocity (\ref
{velocity-homogeneous}) takes the form
\begin{equation}
u^{i}(t)=H(t)x^{i}.  \label{velocity-isotropic}
\end{equation}
If $\{\xi ^{i}\}$ is an initial position of a cosmological particle at $
t=t_{0}$, $x^{i}(t_{0})=\xi ^{i}$, the solution to this initial value
problem gives the equations of motion of a cosmological fluid particle, see
Eq. (\ref{velocity-eqs}), in the following form
\begin{equation}
x^{i}(t)=R(t)\xi ^{i}\hspace{0.4cm}\mathrm{where}\hspace{0.4cm}\frac{1}{R}
\frac{dR}{dt}=H(t),\hspace{0.4cm}R(t_{0})=1.  \label{scale-factor}
\end{equation}
This shows that the only motions of the cosmological fluid compatible with
the Newtonian cosmological principle, that is, with a homogeneous and
isotropic distribution of the fluid, are those of the uniform expansion or
contraction (\ref{scale-factor}) determined by one time-dependent scalar
scale factor $R(t)$ which is always positive, $R(t)\geqslant 0$. This
competes the analysis of the cosmological fluid kinematics due to Newtonian
cosmological principle.

To study the dynamics of the cosmological fluid satisfying the Newtonian
cosmological principle, one should analyze the system of the
Navier-Stokes-Poisson equations (\ref{eq-state2}), (\ref{poisson}), (\ref
{conservation}) and (\ref{navier-stokes}) for the fluid moving in accordance
with Eqs. (\ref{velocity-eqs}) and (\ref{scale-factor}). The equation of
continuity (\ref{conservation}) takes the form
\begin{equation}
\frac{\partial \rho }{\partial t}+3H(t)\rho (t)=\frac{d\rho }{dt}+3H(t)\rho
(t)=0,  \label{continuity-h-i}
\end{equation}
which has a general solution
\begin{equation}
\rho (t)=\frac{\rho (t_{0})}{R^{3}(t)}.  \label{density-h-i}
\end{equation}
The solution (\ref{density-h-i}) admits an obvious interpretation, since if
all linear dimensions in an evolving\ homogeneous and isotropic cosmological
fluid are scaled up by the factor $R(t)$, then all volumes are scaled up by
the factor $R^{3}(t)$ with the correspondent increase or decrease in the
cosmological fluid density.

The Poisson equation for the gravitational potential $\phi =\phi (x^{k},t)$
of a homogeneous and isotropic distribution of the cosmological fluid with
the density $\rho =\rho (t)$ and with the Heckmann-Sch\"{u}cking boundary
condition (\ref{heckmann-schucking}) reads
\begin{equation}
\delta ^{ij}\frac{\partial ^{2}}{\partial x^{i}\partial x^{j}}\phi
(x^{k},t)=4\pi G\rho (t)-\Lambda ,\hspace{0.4cm}E_{ij}(x^{k},t)\rightarrow 0
\hspace{0.4cm}\mathrm{as}\hspace{0.4cm}\left( \delta _{ij}x^{i}x^{j}\right)
^{1/2}\rightarrow \infty .  \label{poisson-h-i}
\end{equation}
This equation has a solution
\begin{equation}
\phi (x^{k},t)=\frac{2}{3}\pi G\rho (t)\delta _{ij}x^{i}x^{j}-\frac{1}{2}
\Lambda \delta _{ij}x^{i}x^{j},\hspace{0.4cm}\mathrm{and}\hspace{0.4cm}\phi
_{,ij}=\frac{4}{3}\pi G\rho (t)\delta _{ij}-\frac{1}{3}\Lambda \delta _{ij}.
\label{potential-h-i}
\end{equation}
The second relation in (\ref{potential-h-i}) shows that the Heckmann-Sch\"{u}
cking boundary condition (\ref{heckmann-schucking}) is satisfied by the
solution and, in fact, the tidal force tensor $E_{ij}$ (\ref
{curvature-newton}) vanishes everywhere
\begin{equation}
E_{ij}=0.  \label{tidal-tensor-zero}
\end{equation}
The physical meaning of the solution (\ref{potential-h-i}) is that the
gravitational potential $\phi (x^{k},t)$ is spherically symmetric and
constant over surfaces $\delta _{ij}x^{i}x^{j}=\mathrm{const}$ for each
moment $t=\mathrm{const}$. Now the Navier-Stokes equation (\ref
{navier-stokes}) takes the form
\begin{equation}
\frac{\partial u^{i}}{\partial t}+u^{j}\frac{\partial u^{i}}{\partial x^{j}}+
\frac{1}{\rho }\delta ^{ij}\frac{\partial p}{\partial x^{j}}=-\delta ^{ij}
\frac{\partial \phi }{\partial x^{j}}\hspace{0.4cm}\rightarrow \hspace{0.4cm}
\left( \frac{dH}{dt}+H^{2}\right) x^{i}=-\frac{4}{3}\pi G\rho (t)x^{i}+\frac{
1}{3}\Lambda x^{i}.  \label{navier-stokes-h-i}
\end{equation}
By expressing the function $H(t)$ and the fluid density $\rho (t)$ in (\ref
{navier-stokes-h-i}) through the scale factor $R(t)$ by (\ref{scale-factor})
and (\ref{density-h-i}) one gets the equation for $R(t)$,
\begin{equation}
R^{2}\frac{d^{2}R}{dt^{2}}+\frac{4}{3}\pi G\rho (t_{0})-\frac{1}{3}\Lambda
R^{3}=0.  \label{eq-r}
\end{equation}
The first integral of this equation (\ref{eq-r}) can be easily found by
integration, to bring the Friedmann equation (\ref{friedmann}) for the scale
factor $R(t)$. Here $-k$ is the integration constant which has meaning of
the total energy of a cosmological fluid particle with a coordinate $R(t)$
moving along its path (\ref{scale-factor}). Indeed the integration of (\ref
{eq-r}) gives
\begin{equation}
-k=\left( \frac{dR}{dt}\right) ^{2}-\frac{8\pi G}{3R}\rho _{0}-\frac{1}{3}
\Lambda R^{2}=K(t)+V(t)  \label{integration-k}
\end{equation}
where $K(t)$ and $V(t)$ are the kinetic and potential energies of the fluid
particle.\hspace{0.4cm}\textbf{QED}
\vspace{0.2cm}

\section{Inhomogeneous and Anisotropic Newtonian Universes}

On the basis of the notion of homogeneous and isotropic Newtonian universe
one can define a notion of an inhomogeneous and/or anisotropic Newtonian
universe.\vspace{0.2cm}

\noindent \textbf{Inhomogeneous and/or anisotropic Newtonian universes}
\textsc{\hspace{0.2cm}}\emph{A Newtonian universe is inhomogeneous and/or
anisotropic if it does not satisfy the Newtonian cosmological principle,
for, at least, an interval of the cosmological evolution time.}\vspace{0.2cm}

As it has been pointed out the Newtonian cosmological principle assumes the
existence of a privileged family of observers who, at each instant of time,
observe the Newtonian universe as homogeneous and isotropic. If a Newtonian
universe is inhomogeneous and/or anisotropic, a free falling observer
comoving with a fluid particle will be measuring deviations in the
velocities, density an pressure for a fluid particle under observation as it
moves, and different values of these quantities for different fluid
particles moving around the observer. From the geometrical point of view
that means that distributions of the evolving cosmological fluid over
3-dimensional Euclidean spaces at each instant of time are different and
therefore the fluid is inhomogeneous and/or anisotropic during its evolution.

\section{The Static Newtonian Universe}

\label{*snu}

The equation (\ref{friedmann}) may be integrated for different values of $
\Lambda $ and $k$ to bring a number of Newtonian cosmological models similar
to the homogeneous and isotropic cosmological models of general relativity
\cite{Trau:1965}, \cite{Bond:1960}, \cite{Heck-Schu:1959}. One of the most
important results of the analysis is that if the Newtonian cosmological
constant vanishes, $\Lambda =0$, then no static Newtonian universes, that
is, either spatially compact or noncompact self-gravitating distributions of
cosmological fluid for which scale factor $R(t)$ (\ref{velocity-isotropic})
and (\ref{scale-factor}) is constant,
\begin{equation}
R(t)=\mathrm{const}\hspace{0.4cm}\rightarrow \hspace{0.4cm}x^{i}(t)=\mathrm{
const}\xi ^{i},\hspace{0.4cm}u^{i}(t)=0,  \label{static}
\end{equation}
can exist \cite{Trau:1965}, \cite{Bond:1960}, \cite{Miln:1934}-\cite
{Heck-Schu:1959}, see also Section \ref{*ibvp}. Therefore, Newtonian
cosmology, as well as general relativistic cosmology, also predicts that our
Universe is evolving in time and there was a time when the evolution began.
If the Newtonian cosmological constant does not vanish, $\Lambda \neq 0$,
then there exists a critical value of the Newtonian cosmological constant $
\Lambda =\Lambda _{c}$ for which the Newtonian universe is static and the
cosmological fluid configuration in this case neither expands, nor
contracts. This Newtonian universe is an analogue of Einstein's static
cosmological model of general relativity.

\begin{corollary}[No static Newtonian universe with $\Lambda =0$]
No static Newtonian universe exists if the cosmological constant vanishes, $
\Lambda =0$. A static Newtonian universe exists for $k>0$ if the Newtonian
cosmological constant $\Lambda $ and the scale factor $R(t)$ have the
critical values, $\Lambda _{c}=k^{3}/16\pi ^{2}G^{2}\rho _{0}^{2}$ and $
R_{c}=4\pi G\rho _{0}/k$.
\end{corollary}

\noindent \textbf{Proof.}\hspace{0.4cm}Let us assume that for a homogeneous
and isotropic Newtonian universe the Newtonian cosmological constant
vanishes, $\Lambda =0$. When the equation of motion (\ref{eq-r}) for the
scale factor $R(t)$ of a fluid particle reads
\begin{equation}
R^{2}\frac{d^{2}R}{dt^{2}}+\frac{4}{3}\pi G\rho (t_{0})=0.
\label{eq-r-static}
\end{equation}
Under the condition of staticity (\ref{static}) Eq. (\ref{eq-r-static})
gives
\begin{equation}
\rho (t_{0})=0  \label{static-density}
\end{equation}
that in contradiction with the definition of a Newtonian cosmology as a
configuration of a self-gravitating cosmological fluid with everywhere
nonvanishing positive fluid density (\ref{mas-positive}) and the positivity
of the mass (I-30) of any fluid region $\Sigma (t)$. Therefore, such a
static fluid configuration is impossible.

Let us now consider the Friedmann equation (\ref{friedmann}) for the scale
factor $R\left( t\right) $. For a static Newtonian universe (\ref{static})
the equation takes the form
\begin{equation}
\frac{8\pi G}{3R}\rho _{0}-k+\frac{1}{3}\Lambda R^{2}=0.
\label{friedmann-static}
\end{equation}
For $k>0$ when the potential energy of a fluid particle is larger than its
kinetic energy (\ref{integration-k}), the critical value of the cosmological
constant, $\Lambda _{c}=k^{3}/16\pi ^{2}G^{2}\rho _{0}^{2}$, ensures that
the function in the right-hand side of Eq. (\ref{friedmann-static}) always
positive for all $R(t)\geqslant 0$ except it vanishes at a double root at $
R(t)=\mathrm{const}=4\pi G\rho _{0}/k=R_{c}$. One can show that for the
cases $k=0$ and $k<0$ this function never vanishes for $R(t)\geqslant 0$.
\hspace{0.4cm}\textbf{QED}
\vspace{0.2cm}

\section{The Newtonian Cosmology versus the General Relativistic Cosmology}

\label{*ncvgrc}

The general relativistic Friedmann equation for the scale factor $R_{\mathrm{
GR}}(t)$ has the same form (\ref{friedmann}) with three differences:

[\emph{a}] the pressure is absent in the case of general relativity;

[\emph{b}] the Newtonian scale factor $R\left( t\right) $ has no physical
dimension while general relativistic scale factor $R_{\mathrm{GR}}(t)$ has
that of time;

[\emph{c}] the constant $k$ which is an integration constant representing
the total energy $-k$ (\ref{integration-k}) in Newtonian cosmology, in
general relativity represents the curvature of 3-dimensional space
orthogonal to the world lines of cosmological fluid particles, $k$ having
one of the value $+1$, $0$ or $-1$ for three different types of
3-dimensional space geometry.

In all other respects the identity of the Newtonian and general-relativistic
Friedmann equation (\ref{friedmann}) is complete \cite{Trau:1965}, \cite
{Bond:1960}, \cite{Miln:1934}, \cite{McCr-Miln:1934}. There are three
physical consequences of this remarkable similarity of fundamental
significance for the construction of a realistic homogeneous and isotropic
cosmological model of our Universe in the frameworks of Newtonian and
general relativistic cosmologies.

\{\emph{a}\} The history of an homogeneous and isotropic Universe as
described in the frameworks of Newtonian and general relativistic
cosmologies is identical for both theories since the Friedmann equation (\ref
{friedmann}) for the variation of the scale factor $dR/dt$ has the same
structure and there is the same set of cosmological models in both
cosmological settings.

\{\emph{b}\} The difference between the Newtonian and general relativistic
cosmologies is governed by the [pressure]/[density] and [gravitational
potential]/[rest mass] ratios. Therefore for the current state of our
Universe where both ratios are always small, general relativity cannot offer
anything radically new. When the propagation of light over the distances
comparable with the radius of the curvature of Universe is to be taken into
consideration, general relativistic cosmology will be more physically
adequate compared with a Newtonian cosmology consideration. However, one
cannot expect any radical differences in both approaches for the scales
characteristic, for example, for galaxies in analysis of galaxy formation,
galaxy structure, etc.

\{\emph{c}\} The ability of general relativity to deal with the cosmological
settings when the [pressure]/[density] and [gravitational potential]/[rest
mass] ratios are not appreciable is not much greater as compared with the
Newtonian cosmology, though the results more easily and directly interpreted
in the framework of general relativity. On the other hand, the mathematical
difficulties\ one encounters while solving cosmological problems in
general-relativistic setting frequently obscure the physical interpretation
of obtained results.

\section{The Navier-Stokes-Poisson Equations in Kinema- tic Quantities}

For analysis of the dynamics of inhomogeneous Newtonian universes, that is,
the evolution of self-gravitating cosmological fluids, it is very useful to
reformulate \cite{Elli:1971} the system of the Navier-Stokes-Poisson
equations (\ref{eq-state2}), (\ref{poisson}), (\ref{conservation}) and (\ref
{navier-stokes}) in terms of the kinematic quantities, see Section I-10.

\subsection{The Evolution Equations}

The first issue in the reformulation of the system is to derive the
evolution equations for the fluid expansion scalar $\theta =\theta (x^{i},t)$
(I-59), the fluid shear tensor $\sigma _{ij}=\sigma _{ij}(x^{k},t)$ (I-58),
and the fluid vorticity tensor $\omega _{ij}=\omega _{ij}(x^{k},t)$ (I-49)
or the fluid vorticity vector $\omega ^{i}=\omega ^{i}(x^{j},t)$\ (I-50)
which must replace the Navier-Stokes equation of motion (\ref{navier-stokes}
) describing the evolution of the fluid velocity vector $
u^{i}=u^{i}(x^{j},t) $ \cite{Elli:1971}.

\begin{theorem}[The evolution equations for the kinematic quantities]
The Navier-Stokes equation (\ref{navier-stokes}) for the velocity vector $
u^{i}$ and the Poisson equation (\ref{poisson}) lead to the following system
of evolution equations for the fluid expansion scalar $\theta (x^{i},t)$,
the fluid shear tensor $\sigma _{ij}(x^{k},t)$ and fluid vorticity vector $
\omega ^{i}(x^{j},t)$:
\end{theorem}

\vspace{0.2cm}
\noindent \textbf{The Raychaudhuri evolution equation for the expansion
scalar }$\theta $
\begin{equation}
\frac{d\theta }{dt}+\frac{1}{3}\theta ^{2}+2(\sigma ^{2}-\omega ^{2})+4\pi
G\rho -\Lambda -\delta ^{ij}A_{i,j}=0,  \label{evol-expansion}
\end{equation}
\vspace{0.2cm}

\noindent \textbf{The propagation equation for the shear tensor }$\sigma
_{ij}$
\begin{equation}
\frac{d\sigma _{ij}}{dt}+\delta ^{kl}\sigma _{ik}\sigma _{lj}+\frac{2}{3}
\theta \sigma _{ij}-\frac{1}{3}\delta _{ij}(2\sigma ^{2}+\omega ^{2}-\delta
^{kl}A_{k,l})+\omega _{i}\omega _{j}+E_{ij}-A_{(i,j)}=0,  \label{evol-shear}
\end{equation}
\vspace{0.2cm}

\noindent \textbf{The propagation equation for the vorticity vector }$\omega
^{i}$
\begin{equation}
\frac{d\omega ^{i}}{dt}+\frac{2}{3}\theta \omega ^{i}-\delta ^{ij}\sigma
_{jk}\omega ^{k}-\frac{1}{2}\varepsilon ^{ijk}A_{j,k}=0.
\label{evol-vorticity}
\end{equation}
\vspace{0.2cm}

\noindent \textbf{Proof.}\hspace{0.4cm}To derive the evolution equations for
the kinematic quantities, one needs to consider the tensor $u_{i,j}$ which
describes the spatial change of the fluid particle's velocity by taking a
spatial derivative of the velocity vector $u_{i}=\delta _{ij}u^{j}$. Due to
the Cauchy decomposition theorem, see Section I-10, the tensor $u_{i,j}$ can
be always represented (I-46) in terms of the fluid expansion scalar $\theta $
(I-59), the fluid shear tensor $\sigma _{ij}$ (I-58) and fluid vorticity
tensor $\omega _{ij}$ (I-49) as\vspace{0.2cm}

\noindent \textbf{The kinematic decomposition of the tensor }$u_{i,j}$
\begin{equation}
u_{i,j}=\sigma _{ij}+\frac{1}{3}\delta _{ij}\theta +\omega _{ij}.
\label{decomposition2}
\end{equation}
\vspace{0.2cm}

\noindent By applying two identities for the commutation of partial derivatives of $
u_{i,j}$ with respect to $t$ and $x^{i}$ \vspace{0.2cm}

\noindent \textbf{The first identity for the tensor}\textsc{\ }$u_{i,j}$
\textsc{\ }
\begin{equation}
u_{i,jt}=u_{i,tj},  \label{identity1}
\end{equation}
\vspace{0.2cm}

\noindent \textbf{The second identity for the tensor }$u_{i,j}$
\begin{equation}
u_{i,jk}=u_{i,kj},  \label{identity2}
\end{equation}
\vspace{0.2cm}

\noindent to a spatial derivative of the Navier-Stokes equation (\ref{navier-stokes})
written with using the total acceleration vector $A^{i}$ (\ref
{acceleration-total-newton}) for a fluid particle
\begin{equation}
u_{i,tj}+\delta ^{kl}u_{i,kj}u_{l}+\delta
^{kl}u_{i,k}u_{l,j}+g_{i,j}-A_{i,j}=0,  \label{navier-stokes-deriv}
\end{equation}
one gets the equation
\begin{equation}
u_{i,jt}+\delta ^{kl}u_{i,jk}u_{l}++g_{i,j}-A_{i,j}=0\hspace{0.4cm}
\rightarrow \hspace{0.4cm}\frac{d}{dt}u_{i,j}+\delta
^{kl}u_{i,k}u_{l,j}+\phi _{,ij}-A_{i,j}=0.
\label{navier-stokes-deriv-ident1-2}
\end{equation}
Making use of the definition of the tidal force tensor $E_{ij}$ (\ref
{curvature-newton}), the Poisson equation (\ref{poisson}) and substituting
the decomposition rule (\ref{decomposition2}) into the spatial derivative of
the Navier-Stokes equation (\ref{navier-stokes-deriv-ident1-2}) with taking
consequently its trace, the symmetric trace-free part and the antisymmetric
part using the definition for the vorticity vector $\omega ^{i}$ (I-50)
brings the the evolution equations for the fluid expansion scalar $\theta $
(\ref{evol-expansion}), the fluid shear tensor $\sigma _{ij}$ (\ref
{evol-shear}), and fluid vorticity vector $\omega ^{i}$ (\ref{evol-vorticity}
).\hspace{0.4cm}\textbf{QED }\vspace{0.2cm}

It should be pointed out here that Eq. (\ref{navier-stokes-deriv-ident1-2})
is the propagation equation for the tensor $u_{i,j}$ along the fluid
particle trajectories. In absence of the fluid pressure when the total fluid
acceleration vector $A^{i}=0$ (\ref{navier-stokes-total}), if one considers
two neighboring free falling fluid particles moving along their trajectories
the equation (\ref{navier-stokes-deriv-ident1-2}) describes a relative
change in the positions of both particles due to action of gravitation
through the second spatial derivative of potential $\phi _{,ij}$. This is
the Newtonian analogue of the geodesic deviation equation in general
relativity \cite{Pira:1965} and the quantity $\phi _{,ij}$ or the tidal
force tensor $E_{ij}$ (\ref{curvature-newton}), is the Newtonian analogue of
the space-time curvature tensor of General Relativity \cite{Trau:1965}, \cite
{Pira:1965}.

\subsection{The Constraint Equations}

The identities (\ref{identity1}) and (\ref{identity2}) for the tensor $
u_{i,j}$ are Newtonian analogues of the so-called Ricci identities in
general relativity, which relate the second derivatives of an arbitrary
vector field with the space-time curvature tensor \cite{Elli:1971}.

Due to the second identity (\ref{identity2}) for the tensor $u_{i,j}$ there
is also another set of equations which put constraints on the kinematic
quantities \cite{Elli:1971}.

\begin{theorem}[The constraints on the kinematic quantities]
The second identity\linebreak\ for the tensor $u_{i,j}$ (\ref{identity2})
puts the following constraints on the fluid expansion scalar $\theta
(x^{i},t)$, the fluid shear tensor $\sigma _{ij}(x^{k},t)$ and fluid
vorticity vector $\omega ^{i}(x^{j},t)$:
\end{theorem}

\vspace{0.2cm}
\noindent \textbf{The first constraint equation}
\begin{equation}
\delta ^{jk}(\sigma _{ij,k}-\omega _{ij,k})-\frac{2}{3}\theta _{,j}=0,
\label{constraint-1}
\end{equation}
\vspace{0.2cm}

\noindent \textbf{The second constraint equation}\textsc{\ }
\begin{equation}
\omega _{,i}^{i}=0,  \label{constraint-2}
\end{equation}
\vspace{0.2cm}

\noindent \textbf{The third constraint equation}
\begin{equation}
\delta ^{j(i}(\sigma _{jk,l}+\omega _{jk,l})\varepsilon ^{m)kl}=0.
\label{constraint-3}
\end{equation}
\vspace{0.2cm}

\noindent \textbf{Proof.}\hspace{0.4cm}The first constraint (\ref
{constraint-1}) follows from taking a trace of the second identity (\ref
{identity2}) for the tensor $u_{i,j}$ with respect to indices $i$ and $j$.
The second constraint (\ref{constraint-2}) follows from the fact that the
total antisymmetrization of (\ref{identity2}) in its indices gives
\begin{equation}
u_{[i,jk]}=0  \label{identity2-anti}
\end{equation}
which provides (\ref{constraint-2}) after total contraction with the
Levi-Civita symbol $\varepsilon ^{ijk}$. The third constraint (\ref
{constraint-3}) follows after the contraction of the second identity for the
tensor $u_{i,j}$ (\ref{identity2}) with $\varepsilon ^{mjk}$ and
symmetrization of the two remaining free indices
\begin{equation}
\delta ^{i(n}\varepsilon ^{m)jk}\left( u_{i,jk}-u_{i,kj}\right) =0.
\label{identity2-sym}
\end{equation}
\textbf{QED}\vspace{0.2cm}

It should be noted here that the trace of the constraint (\ref{constraint-3}
) vanishes identically due to the constraint (\ref{constraint-2}).

\subsection{The Integrability Conditions}

There are three more constraints on the kinematic quantities following from
the first and the second identities (\ref{identity1}) and (\ref{identity2})
for the tensor $u_{i,j}$ as their integrability conditions.

\begin{theorem}[The intergrability conditions for the kinematic quantities]
There\linebreak\ is the following set of the integrability conditions of the
first identity (\ref{identity1}) and the second identity (\ref{identity2})
for the tensor $u_{i,j}$. No more integrability conditions exist.
\end{theorem}

\vspace{0.2cm}
\noindent \textbf{The first integrability condition}\textsc{\ }
\begin{equation}
E_{i,j}^{j}=\frac{8\pi G}{3}\rho _{,i}  \label{integrability-1}
\end{equation}
\vspace{0.2cm}

\noindent \textbf{The second integrability condition}
\begin{equation}
E_{k,j}^{(i}\varepsilon ^{l)kj}=0,  \label{integrability-2}
\end{equation}
\vspace{0.2cm}

\noindent \textbf{The third integrability condition}
\begin{equation}
\sigma _{\lbrack k,l]}^{[i,j]}+\frac{2}{3}\delta _{\lbrack k}^{[i}\theta
_{,l]}^{,j]}=0.  \label{integrability-3}
\end{equation}
\vspace{0.2cm}

\noindent \textbf{Proof.}\hspace{0.4cm}The derivation of the integrability
conditions is straightforward. The first and the second intergrability
conditions follow from taking a spatial derivative of the first identity
(\ref{identity1}) and the requirement $u_{i,t[jk]}=0$ which brings an
integrability condition
\begin{equation}
u_{i,jtk}-u_{i,ktj}-u_{i,t[jk]}=0\hspace{0.4cm}\rightarrow \hspace{0.4cm}
u_{i,jtk}-u_{i,ktj}=0.  \label{integrability-1-2}
\end{equation}
Taking the trace of (\ref{integrability-1-2}) with respect to indices $i$
and $j$ with using the evolution equations and the Poisson equation (\ref
{poisson}) gives the first integrability condition (\ref{integrability-1}),
while the antisymmetrization of (\ref{integrability-1-2}) with respect to
indices $i$ and $j$ gives the second integrability condition (\ref
{integrability-2}). No more constraints on the kinematic quantities can be
found from the integrability\ condition (\ref{integrability-1-2}). The third
intergrability conditions follows from taking a spatial derivative of the
second identity (\ref{identity2}) and the requirement $u_{i,j[kl]}=0$ which
brings an integrability condition
\begin{equation}
u_{i,kjl}-u_{i,ljk}-u_{i,j[kl]}=0\hspace{0.4cm}\rightarrow \hspace{0.4cm}
u_{i,kjl}-u_{i,ljk}=0.  \label{integrability-3-0}
\end{equation}
Taking the antisymmetrization of (\ref{integrability-3-0}) with respect to
indices $i$ and $j$ gives the third integrability condition (\ref
{integrability-3}). No more constraints on the kinematic quantities can be
found from the integrability\ condition (\ref{integrability-3-0}).

It is easy to show that calculations of the integrability conditions for
(\ref{integrability-1-2}) and (\ref{integrability-3-0}) does not bring any
new constraints since they are satisfied identically.\hspace{0.4cm}\textbf{
QED}\vspace{0.2cm}

It should be noted here that the trace of the integrability condition (\ref
{integrability-2}) vanishes identically because the tensor $E_{ij}$ is
symmetric.

The integrability conditions (\ref{integrability-1}), (\ref{integrability-2}
) and (\ref{integrability-3}) are analogues of the Bianchi identities for
the space-time curvature tensor of general relativity \cite{Elli:1971} that
can be seen from the structure of the integrability conditions (\ref
{integrability-1-2}) and (\ref{integrability-3-0}) which are just a cyclic
combination for the third derivatives of the fluid velocity $u^{i}$.
Therefore taking further integrability conditions results in expressions
identically vanishing in a 3-dimensional space for they involve the
antisymmetrization of more than three indices.

\section{The Structure of the System of Equations}

While the Poisson equation (\ref{poisson}) keeps the same form in this
reformulation of the Navier-Stokes-Poisson equations in terms of the
kinematic quantities, the equation of continuity (\ref{conservation}) takes
the following form.\vspace{0.2cm}

\noindent \textbf{The equation of continuity as the evolution equation for
the fluid density}\textsc{\ }$\rho $
\begin{equation}
\frac{d\rho }{dt}+\rho \theta =0.  \label{evol-density}
\end{equation}
\vspace{0.2cm}

Thus, the system of the Navier-Stokes-Poisson equations for the fluid
expansion scalar $\theta (x^{i},t)$, the fluid shear tensor $\sigma
_{ij}(x^{k},t)$ and fluid vorticity vector $\omega ^{i}(x^{j},t)$ includes
the evolution equations (\ref{evol-expansion})-(\ref{evol-vorticity}), the
constraints (\ref{constraint-1})-(\ref{constraint-3}), the integrability
conditions (\ref{integrability-1})-(\ref{integrability-3}), the equation of
continuity (\ref{evol-density}), the Poisson equation (\ref{poisson}) and
(\ref{eq-state2}), to replace the Navier-Stokes-Poisson equations (\ref
{eq-state2}), (\ref{identity}), (\ref{poisson}), (\ref{conservation}) and
(\ref{navier-stokes}). Upon having determined the kinematic quantities by
solving the above system, one must solve the nonhomogeneous equations (\ref
{decomposition2}) for the fluid velocity $u^{i}(x^{j},t)$ together with the
identities (\ref{identity1}) and (\ref{identity2}). Finally, to determine
the equation of motion of fluid particles $x^{i}=x^{i}(\xi ^{j},t)$, or $\xi
^{j}=\xi ^{j}(x^{i},t)$, (I-1)-(I-4) from the fluid velocity $u^{i}(x^{j},t)$
the initial value problem (\ref{velocity-eqs}) has to be solved.

This system of equations for the kinematic quantities and the
Navier-Stokes-Poisson equations are usually considered to be equivalent \cite
{Elli:1971}. However, this has not been shown as yet, strictly speaking. As
it has been noted above, the Navier-Stokes-Poisson equations is a system of
a mixed hyperbolic-elliptic type for seven first order nonlinear partial
differential equations for seven unknowns $\rho $, $u^{i}$ and $g_{i}$, or
it can be taken as a system of five first and second order equations (\ref
{poisson}), (\ref{conservation}) and (\ref{navier-stokes}) for five unknowns
$\rho $, $u^{i}$ and $\phi $ to be supplemented by the corresponding initial
and boundary conditions for the unknowns and an equation of state (\ref
{eq-state2}). It allows application of some techniques for analyzing the
questions of existence and uniqueness, at least, for a class of boundary
and/or initial conditions, see Section \ref{*nspe}. Once reformulated in
terms of kinematic quantities, the system of nonlinear equations for a
self-gravitating fluid changes its structure and it is does not fit into the
standard classification of the partial differential equations. Four main
issues appear to be important in consideration of the Navier-Stokes-Poisson
equations written in terms of kinematic quantities:

(A) The system is not closed since the evolution equation for the shear
tensor (\ref{evol-shear}) contains the tidal force tensor $E_{ij}$ (\ref
{curvature-newton}) which does not have any evolution equation for itself.
This is a reflection of the noncloseness of the Navier-Stokes-Poisson
equations (\ref{eq-state2}), (\ref{poisson}), (\ref{conservation}) and (\ref
{navier-stokes}) with respect to boundary conditions and its initial
boundary value problem, see Section \ref{*ibvp}. Upon reformulation in terms
of kinematic quantities this problem becomes, however, critical for the
Navier-Stokes-Poisson equations. Indeed, the system (\ref{poisson}), (\ref
{conservation}) and (\ref{navier-stokes}) can be supplemented by some
boundary conditions, for example, the Heckmann-Sch\"{u}cking boundary
conditions (\ref{heckmann-schucking}) or (\ref{heckmann-schucking2}), to
allow its solution without adding any other equations and an explicit change
in the structure of the system of equations (\ref{eq-state2}), (\ref{poisson}
), (\ref{conservation}) and (\ref{navier-stokes}). On the contrary, the
system (\ref{eq-state2}), (\ref{poisson}), (\ref{navier-stokes-total}), (\ref
{evol-expansion})-(\ref{identity2}), (\ref{constraint-1})-(\ref{constraint-3}
), (\ref{integrability-1})-(\ref{integrability-3}) and (\ref{evol-density})
must be explicitly supplemented by an evolution equation for the tidal force
tensor $E_{ij}$ (\ref{curvature-newton}) in order to allow its solution.
Thus, the Navier-Stokes-Poisson equations in terms of kinematic quantities
must admit a generalization of Newtonian gravity in terms of an evolution
equation for the $E_{ij}$ in addition to the Poisson equation (\ref{poisson}
) which then acquires a status of a constraint equation. As it has been
pointed out in Section \ref{*ibvp}, there are a number of approaches known
to modify the Newtonian gravity satisfying the Poisson equation, see \cite
{Szek-Rain:2000}, \cite{Bert-Hami:1994}, \cite{Elli-Duns:1994} and
references therein. One possibility is to accept the so-called local tidal
approximation \cite{Bert-Hami:1994}\vspace{0.2cm}

\noindent \textbf{The evolution equations for the tidal force tensor }$
E_{ij} $
\begin{equation}
\frac{dE_{ij}}{dt}+\theta E_{ij}-\delta _{ij}\delta ^{kn}\delta ^{lm}\sigma
_{kl}E_{nm}-3\delta ^{kl}\sigma _{k(i}E_{j)l}-\delta ^{kl}\omega
_{k(i}E_{j)l}+4\pi G\rho \sigma _{ij}=0  \label{evol-tidal}
\end{equation}
\vspace{0.2cm}

\noindent which is a generalization of the Zel'dovich approximation \cite{Zel':1970},
\cite{Zel'-Novi:1974} by taking into account the second spatial derivatives
in the equation of motion for a fluid particle \cite{Hui-Bert:1996}. With
such an evolution equation (\ref{evol-tidal}) for the tidal force tensor $
E_{ij}$ (\ref{curvature-newton}) added, the system of the
Navier-Stokes-Poisson equations in kinematic quantities (\ref{eq-state2}),
(\ref{poisson}), (\ref{navier-stokes-total}), (\ref{evol-expansion})-(\ref
{identity2}), (\ref{constraint-1})-(\ref{constraint-3}), (\ref
{integrability-1})-(\ref{integrability-3}) and (\ref{evol-density}) becomes
closed. Then one can look for a solution of the system with a suitable set
of initial and boundary conditions characteristic for the problem under
study. One can assume another evolution equation for the tidal force tensor $
E_{ij}$ which may be dictated by a particular Newtonian cosmological model.

(B) The system is overdetermined since it has been derived by
differentiation of the Navier-Stokes-Poisson equations (\ref{eq-state2}),
(\ref{poisson}), (\ref{conservation}) and (\ref{navier-stokes}). It now
contains higher derivatives of the fluid velocity through the expansion
scalar $\theta =\theta (x^{k},t)$ (I-59), the shear tensor $\sigma
_{ij}=\sigma _{ij}(x^{k},t)$ (I-58) and the vorticity vector $\omega
^{i}=\omega ^{i}(x^{k},t)$ (I-50) and these kinematic became now additional
unknowns to the velocity field $u^{i}=u^{i}(x^{j},t)$, the fluid density
scalar $\rho =\rho (x^{i},t)$ and the gravitational potential scalar $\phi
=\phi (x^{i},t)$. There are 14 unknowns, namely, 3 components of the fluid
velocity $u^{i}$, 1 function of the density $\rho $, 1 function of the
gravitational potential $\phi $, 1 function of the expansion $\theta $, 5
components of the trace-free symmetric shear tensor $\sigma _{ij}$ and 3
components of the fluid vorticity vector $\omega ^{i}$. All together they
should satisfy 69 equations, namely, 9 evolution equations (\ref
{evol-expansion})-(\ref{evol-vorticity}), the equation of continuity (\ref
{evol-density}), the equation of state (\ref{eq-state2}), the Poisson
equation (\ref{poisson}), 5 evolution equations (\ref{evol-tidal}), 9
constraints including 3 constraints (\ref{constraint-1}), 1 constraint (\ref
{constraint-2}), 5 constraints (\ref{constraint-3}), 16 integrability
conditions including 3 integrability conditions (\ref{integrability-1}), 5
integrability conditions (\ref{integrability-2}) and 8 integrability
conditions (\ref{integrability-3}) and 9 constraints for the velocity (\ref
{decomposition2}) which satisfy 18 integrability conditions (\ref{identity1}
) and (\ref{identity2}).

(C) Even after the reformulation in terms of the kinematic quantities of the
fluid expansion scalar $\theta =\theta (x^{i},t)$ (I-59), the fluid shear
tensor $\sigma _{ij}=\sigma _{ij}(x^{k},t)$ (I-58), and the fluid vorticity
tensor $\omega _{ij}=\omega _{ij}(x^{k},t)$ (I-49) or the fluid vorticity
vector $\omega ^{i}=\omega ^{i}(x^{j},t)$\ (I-50) the system of the
Navier-Stokes-Poisson equations (\ref{eq-state2}), (\ref{poisson}), (\ref
{navier-stokes-total}), (\ref{evol-expansion})-(\ref{identity2}), (\ref
{constraint-1})-(\ref{constraint-3}), (\ref{integrability-1})-(\ref
{integrability-3}), (\ref{evol-density}), (\ref{curvature-newton}) and (\ref
{evol-tidal}) do contain the fluid velocity $u^{i}(x^{j},t)$ explicitly.
Indeed, the material derivatives in the evolution equations (\ref
{evol-expansion})-(\ref{evol-vorticity}) have the inertial term $
u_{i,j}u^{j} $ involving $u^{i}(x^{j},t)$. That means from the mathematical
point of view that the information stored in the spatial derivatives of the
fluid velocity $u_{i,j}(x^{k},t)$ through the decomposition (\ref
{decomposition2}) is not enough to fully specify the fluid velocity field $
u^{i}(x^{j},t)$ in terms of the kinematic quantities. Therefore, if only the
kinematic quantities $\theta (x^{i},t)$, $\sigma _{ij}(x^{k},t)$ and $\omega
^{i}(x^{j},t)$ are considered to be the unknowns, the system of the
Navier-Stokes-Poisson equations is actually a system of integro-differential
equations.

(D) The space of solutions of the system (\ref{eq-state2}), (\ref{poisson}),
(\ref{navier-stokes-total}), (\ref{evol-expansion})-(\ref{identity2}), (\ref
{constraint-1})-(\ref{constraint-3}), (\ref{integrability-1})-(\ref
{integrability-3}), (\ref{evol-density}), (\ref{curvature-newton}) and (\ref
{evol-tidal}) is bigger than that of the Navier-Stokes-Poisson equations
(\ref{eq-state2}), (\ref{poisson}), (\ref{conservation}) and (\ref
{navier-stokes}). This is reflected in the presence, first of all, of an
evolution equation for the tidal force tensor (\ref{evol-tidal}) which does
not take place in the original system. Secondly, there are constraints and
integrability conditions which include even higher derivatives of the tensor
$u_{i,j}$. Since the decomposition law (\ref{decomposition2}) is a Pfaffian
system when it is considered as a system of nonhomogeneous partial
differential equations for the 3 unknown components of velocity $u_{i}$, due
to the Frobenius theorem \cite{Koba-Nomi:1963}, \cite{BCGGG:1991} the
integrability conditions (\ref{identity1}) and (\ref{identity2}) are
necessary and sufficient for the existence of a solution locally. The
problem, however, is that the identities (\ref{identity1}) and (\ref
{identity2}) taken as a system of nonhomogeneous partial differential
equations for the tensor $u_{i,j}$ form a system of partial differential
equations which is not Pfaffian. In such a case, their integrability
conditions (\ref{integrability-1})-(\ref{integrability-3}) are only
sufficient conditions for integrability of (\ref{identity1}) and (\ref
{identity2}) in general (see, for instance, \cite{BCGGG:1991} and references
therein). One may expect they become also necessary for some cosmological
fluid configurations. When and under which assumptions about the physics of
a self-gravitating fluid, and finally, for which class(es) of Newtonian
cosmologies this can occur definitely deserves a careful study. It should be
pointed out here that this problem of the integrability of the
Navier-Stokes-Poisson equations in kinematic quantities is affected also by
the structure of the evolution equations for the tidal force tensor $E_{ij}$
(\ref{evol-tidal}), or another equation of this kind, which has been added
to close the system.

(E) The dynamics of the self-gravitating fluid governed by the system of
Navier-Stokes-Poisson equations in terms of kinematic quantities (\ref
{eq-state2}), (\ref{poisson}), (\ref{navier-stokes-total}), (\ref
{evol-expansion})-(\ref{identity2}), (\ref{constraint-1})-(\ref{constraint-3}
), (\ref{integrability-1})-(\ref{integrability-3}), (\ref{evol-density}),
(\ref{curvature-newton}) and (\ref{evol-tidal}) is not, strictly speaking,
due to Newtonian gravity of the cosmological fluid, but rather due to its
generalization in accordance with the evolution equation (\ref{evol-tidal})
for the gravitational potential through the tidal force tensor $E_{ij}$ (\ref
{curvature-newton}). The universes corresponding to such dynamically
evolving cosmological fluid configurations under this modified Newtonian
gravitational force will be still called Newtonian, since if the Newtonian
cosmological principle is satisfied, the tidal force tensor should vanish
asymptotically (\ref{heckmann-schucking}) and (\ref{heckmann-schucking2}) on
the largest cosmological scales and the Newtonian universes homogeneous and
isotropic (\ref{friedmann}) on such large scales are governed then by the
Newtonian gravitational potential satisfying the Poisson equation (\ref
{poisson}). If a Newtonian universe is inhomogeneous and/or anisotropic the
full system of the Navier-Stokes-Poisson equations in terms of kinematic
quantities must be taken into consideration.

Though the system of the Navier-Stokes-Poisson equations in terms of
kinematic quantities look a bit complicated as compared with the original
Navier-Stokes-Poisson equations, the former has proved to be mathematically
efficient and physically adequate in cosmological studies, in particular,
where some simplifying assumptions can be adopted (see \cite{Elli:1971} and
recent applications \cite{Sota-etal:1998}-\cite{Barr-Gotz:1989} and
references therein).

\section{The Raychaudhuri Equation in a Homogeneous and Isotropic Newtonian
Universe}

The evolution equation for the expansion scalar (\ref{evol-expansion}) is
known to have the very similar form to the Raychaudhuri equation for the
expansion scalar of the 4-velocity of the cosmological fluid in general
relativity \cite{Rayc:1955}, \cite{Hawk:1966} with the only different term
with pressure involved in the general relativistic case. This equation is
very significant in both Newtonian and general relativistic cosmologies
since the expansion scalar of the evolving cosmological fluid serves as its
fundamental parameter. For an homogeneous and isotropic Newtonian universe
it is the only dynamical fluid characteristic. One can show in this case
that the Raychaudhuri equation (\ref{evol-expansion}) is equivalent for the
Friedmann equation (\ref{friedmann}), and the expansion scalar is then
directly expressible in terms of scale factor \cite{Heck-Schu:1955}, \cite
{Heck-Schu:1956}, \cite{Heck-Schu:1959}.

\begin{theorem}[The Homogeneous and Isotropic Raychaudhuri Equation]
The ev\-olution of a Newtonian universe satisfying the Newtonian
cosmological principle is governed by the Raychauhuri equation
\begin{equation}
\frac{d\theta }{dt}+\frac{1}{3}\theta ^{2}+4\pi G\rho -\Lambda =0
\label{Rauchaudhuri-h-i}
\end{equation}
for the cosmological fluid expansion scalar $\theta =\theta (t)$, while the
shear tensor $\sigma _{ij}(x^{k},t)$, vorticity vector $\omega ^{i}(x^{j},t)$
and the tidal force tensor $E_{ij}(x^{k},t)$ vanish
\begin{equation}
\sigma _{ij}=0,\hspace{0.4cm}\omega ^{i}=0,\hspace{0.4cm}E_{ij}=0.
\label{shear-vorticity-tidal-zero}
\end{equation}
Then the Raychauhuri equation is equivalent to the Friedmann equation (\ref
{friedmann}) for the fluid scale factor $R=R(t)$ with
\begin{equation}
\theta =\frac{3}{R}\frac{dR}{dt}.  \label{expansion-scale}
\end{equation}
\end{theorem}

\noindent \textbf{Proof.}\hspace{0.4cm}Let us consider a Newtonian universe
filled with a cosmological fluid satisfying the Newtonian cosmological
principle and governed by the system of Navier-Stokes-Poisson equations in
terms of kinematic quantities (\ref{eq-state2}), (\ref{poisson}), (\ref
{navier-stokes-total}), (\ref{evol-expansion})-(\ref{identity2}), (\ref
{constraint-1})-(\ref{constraint-3}), (\ref{integrability-1})-(\ref
{integrability-3}), (\ref{evol-density}), (\ref{curvature-newton}) and (\ref
{evol-tidal}). The same analysis as in the case of an homogeneous and
isotropic Newtonian universe satisfying the Newtonian cosmological
principle, Section \ref{*ncp}, shows that the fluid density $\rho (x^{k},t)$
and fluid pressure $p(x^{k},t)$ are functions of absolute time $t$ only and
the velocity of the cosmological fluid $u^{i}(x^{k},t)$ has the following
form,
\begin{equation}
\rho =\rho (t),\hspace{0.4cm}p=p(t),\hspace{0.4cm}u^{i}(t)=H(t)x^{i}.
\label{h-i-universe}
\end{equation}
Calculation of the tensor $u_{i,j}$ and the kinematic quantities for the
fluid gives
\begin{equation}
u_{i,j}=H(t)\delta _{ij},\hspace{0.4cm}\theta =H(t),\hspace{0.4cm}\sigma
_{ij}=0,\hspace{0.4cm}\omega ^{i}=0.  \label{kinematic-h-i}
\end{equation}
The kinematic quantities (\ref{kinematic-h-i}) together with a solution to
the equation of continuity (\ref{density-h-i}) satisfy the equations (\ref
{eq-state2}), (\ref{evol-vorticity})-(\ref{identity2}), (\ref{constraint-1}
)-(\ref{constraint-3}), (\ref{integrability-3}) and (\ref{evol-density})
from the system of Navier-Stokes-Poisson equations in terms of kinematic
quantities. The evolution equation (\ref{evol-shear}) for the shear tensor
results in everywhere vanishing tidal force tensor $E_{ij}$ for the fluid
(\ref{tidal-tensor-zero}) which satisfies the evolution equation for the
tidal force tensor (\ref{evol-tidal}). The vanishing tensor $E_{ij}$ (\ref
{tidal-tensor-zero}) is consistent with the solution (\ref{potential-h-i})
of the Poisson equation (\ref{poisson}). Then the Raychaudhuri evolution
equation for the expansion scalar $\theta =H(t)$ is the only remaining
equation which takes the form (\ref{Rauchaudhuri-h-i}). By taking into
account the definition of the scale factor $R(t)$ of the cosmological fluid
(\ref{scale-factor}), see also the representation of the expansion scalar
(I-61), one can easily derive the evolution equation (\ref{eq-r}) for the
scale factor from the Raychaudhuri equation (\ref{Rauchaudhuri-h-i}). The
first integral of (\ref{eq-r}) leads the Friedmann equation (\ref{friedmann}
) for the scale factor $R(t)$.\hspace{0.4cm}\textbf{QED\vspace{0.2cm}}

This result shows, in particular, that for the case of homogeneous and
isotropic Newtonian universes the system of Navier-Stokes-Poisson equations
in terms of kinematic quantities (\ref{eq-state2}), (\ref{poisson}), (\ref
{navier-stokes-total}), (\ref{evol-expansion})-(\ref{identity2}), (\ref
{constraint-1})-(\ref{constraint-3}), (\ref{integrability-1})-(\ref
{integrability-3}), (\ref{evol-density}), (\ref{curvature-newton}) and (\ref
{evol-tidal}) is consistent.

\section*{Acknowledgments}

I would like to thank Alan Coley for his hospitality in Dalhousie
University. The work has been supported in part by the Swiss National
Science Foundation, Grant 7BYPJ065731.

\end{document}